\documentclass[12pt,floatfix,aps,amsmath,amssymb,preprintnumbers,nofootinbib, notitlepage, prepint]{revtex4-1}
\pdfoutput=1
\usepackage[]{graphicx}
\usepackage{color}
\usepackage{hyperref}					
\usepackage{mathtools}					
\usepackage{slashed}				
\usepackage{algorithm,algorithmic}	
\usepackage{dcolumn}	
\usepackage{bm}		
\usepackage{bbm}
\usepackage{pxfonts}
\usepackage{subfigure}

\usepackage{hyperref}
\usepackage{url}
	
\usepackage[vcentermath]{youngtab}
\usepackage{feynmp}
\def\beq{\begin{equation}}
\def\eeq{\end{equation}}
\makeatletter
\def\endfmffile{%
  \fmfcmd{\p@rcent\space the end.^^J%
          end.^^J%
          endinput;}%
  \if@fmfio
    \immediate\closeout\@outfmf
  \fi
  \IfFileExists{\thefmffile.mp}{\immediate\write18{mpost \thefmffile}}{}
  \let\thefmffile\relax
}
\makeatother

\usepackage{bbold}

\newcommand{\be}{\begin{equation}}
\newcommand{\ee}{\end{equation}}
\newcommand{\bea}{\begin{eqnarray}} 
\newcommand{\eea}{\end{eqnarray}}

\newcommand{\bmp}{\noindent\begin{minipage}{16cm}}
\newcommand{\emp}{\end{minipage}\vskip 7mm} 
\def\lsim{\mathrel{\raise.3ex\hbox{$<$\kern-.75em\lower1ex\hbox{$\sim$}}}}
\def\gsim{\mathrel{\raise.3ex\hbox{$>$\kern-.75em\lower1ex\hbox{$\sim$}}}}

\newcommand{\intron}[1]{}

\definecolor{rossoCP3}{cmyk}{0,.88,.77,.40}
\definecolor{light_gray}{rgb}{0.8,0.8,0.8}

\long\def\del #1 \enddel { }

\begin{document}

\title{ \Large  \color{rossoCP3} Quantum Critical Behaviour \\of \\  Semi-Simple Gauge Theories  }
\author{Jacob Kamuk Esbensen$^{\color{rossoCP3}}$}
\email{esbensen@cp3.dias.sdu.dk} 
\author{Thomas A. Ryttov$^{\color{rossoCP3}}$}
\email{ryttov@cp3.dias.sdu.dk} 
\author{Francesco  Sannino$^{\color{rossoCP3}}$}
\email{sannino@cp3.dias.sdu.dk} 
\affiliation{\mbox{
$^{\color{rossoCP3}}$
$CP^3$-Origins \& the Danish IAS,
Univ. of Southern Denmark, Campusvej 55, DK-5230 Odense}}

\begin{abstract}
We study the perturbative phase diagram of semi-simple fermionic gauge theories resembling the Standard Model.  We investigate an $SU(N)$ gauge theory with $M$ Dirac flavors where we gauge first an $SU(M)_L$ and then an $SU(2)_L \subset SU(M)_L$ of the original global symmetry $SU(M)_L\times SU(M)_R \times U(1) $ of the theory. To avoid gauge anomalies we add lepton-like particles. At the two-loops level an intriguing phase diagram appears. 
We uncover phases in which one, two or three fixed points exist and discuss the associated flows of the coupling constants. We discover a phase featuring complete asymptotic freedom and simultaneously an interacting infrared fixed point in both couplings. The analysis further reveals special renormalisation group trajectories along which one coupling displays asymptotic freedom and the other asymptotic safety, while both flowing in the infrared to an interacting fixed point. These are \emph{safety free} trajectories. We briefly sketch out possible phenomenological implications, among which an independent way to generate near-conformal dynamics a l\'a walking is investigated. 
 \vskip.7cm
{\noindent \footnotesize Preprint: CP3-Origins-2015-051, DNRF90 \& DIAS-2015-51}
\end{abstract}

\maketitle


\newpage

\section{Introduction}

The Standard Model is an example of a semi-simple gauge theory with two non-abelian and a single abelian gauge group. It is in addition a non-supersymmetric gauge-Yukawa theory that contains spin-0, spin-$\frac{1}{2}$ as well as spin-1 particles. Whereas the literature contains investigations of the phase diagram of semi-simple and simple supersymmetric gauge theories \cite{Intriligator:2005aw,Seiberg:1994pq,Ryttov:2007sr}, vector-like fermionic gauge theories \cite{Caswell:1974gg,Banks:1981nn,Dietrich:2006cm,Sannino:2004qp,Pica:2010xq,Ryttov:2010iz,Shrock:2015owa,Shrock:2013pya,Shrock:2013ca}, chiral gauge theories \cite{Shi:2015fna,Shi:2015baa,Shi:2014yxa}, Yukawa theories \cite{Mojaza:2011rw,Grinstein:2011dq,Antipin:2011aa,Molgaard:2014mqa,Krog:2015bca} and scalar theories \cite{Shrock:2014zca} only very little has been done concerning the more general class of theories in which the Standard Model falls \cite{Antipin:2013sga}. 

Recently however much interest has been given to gauge-Yukawa theories  featuring a single gauge group \cite{Litim:2014uca}. The reason for such an interest is that this theory was found to flow to a nontrivial ultraviolet stable fixed point in a completely controllable manner \cite{Litim:2014uca,Litim:2015iea}. The matter sector contains a set of $N_f$ Dirac fermions and $N_f\times N_f$ scalar mesons. 
The result shows that no additional symmetry principles, such as space-time supersymmetry \cite{Bagger:1990qh}, are required to ensure well-defined and predictive ultraviolet theories. The ultraviolet fixed point arises dynamically through  renormalizable interactions between non-Abelian gauge fields, fermions, and scalars, and in a regime where asymptotic freedom is absent.
Furthermore the dangerous growth of the gauge coupling towards the ultraviolet is countered by Yukawa interactions, while the Yukawa and scalar couplings are tamed by the fluctuations of  gauge and fermion fields. This has led to the discovery of  {\it complete asymptotic safety}, meaning an interacting ultraviolet  fixed point in all couplings \cite{Litim:2014uca}. This phenomenon is quite distinct  from the conventional setup of {\it complete asymptotic freedom} \cite{Gross:1973ju,Cheng:1973nv,Callaway:1988ya}, where the ultraviolet dynamics of Yukawa and scalar interactions is tamed by asymptotically free gauge fields; see \cite{Holdom:2014hla,Giudice:2014tma} for recent studies.
It is, indeed, very important that it is only via the combined analysis of the gauge, Yukawa and scalar self couplings that one observes the existence of the ultraviolet fixed point. For instance it is not a feature of the gauge theory with either pure fermionic or scalar matter. Neither does the ultraviolet fixed point exist for the supersymmetrized version \cite{Intriligator:2015xxa,Martin:2000cr}. It is also straightforward to engineer QCD-like IR behaviour including confinement and chiral symmetry breaking. In practice one needs to decouple, at some intermediate energies, the unwanted fermions by adding mass terms or via spontaneous symmetry breaking in such a way that at lower energies the running of the gauge coupling mimics QCD \cite{Sannino:2015sel}. Tantalising indications that ultraviolet interacting fixed point may exist nonperturbatively, and without the need of elementary scalars, appeared in \cite{Pica:2010xq}, and they were further explored in  \cite{Shrock:2013cca,Litim:2014uca}. Nonperturbative techniques are needed to establish the existence of such a fixed point when the number of colors and flavours is taken to be three and the number of UV light flavours is large but finite.  Asymptotic safety was originally introduced by Weinberg \cite{Weinberg:1980gg}  to address quantum aspects of  gravity  \cite{Litim:2011cp,Litim:2006dx,Niedermaier:2006ns,Niedermaier:2006wt,
Percacci:2007sz,Litim:2008tt,Reuter:2012id,Dona:2015tnf}.

These observations should make it clear that further studies of gauge-Yukawa theories are to be carried out. Here we take one step further and study non-supersymmetric semi-simple gauge theories with fermionic matter.  This is the logical next step given that a trait d'union of the majority of the extensions of the Standard Model (and the Standard Model itself) is the presence of multiple gauge couplings. 
Furthermore the theories we study are chosen such as to resemble the Standard Model in a most natural way and candidate theories of fermion mass generation \cite{Raby:1979my,Kaplan:1991dc,Barnard:2013zea,Ferretti:2013kya,Cacciapaglia:2015yra} in elementary \cite{Alanne:2014kea,Gertov:2015xma} or composite extensions of the Standard Model \cite{Weinberg:1975gm,Susskind:1978ms,Kaplan:1983fs,Kaplan:1983sm}. 

To construct our case study we begin with an $SU(N)$ gauge theory with  $M$ Dirac fermions in the fundamental representation. We then gauge a subgroup of the global symmetry $SU(M)_L \times SU(M)_R \times U(1)$ and study the flow of both gauge couplings at  the two loop level. 

Since we gauge a subgroup of the global symmetry we have to worry about potential anomalies. In order for our theory to be consistent we therefore add a set of lepton-like fermions.  We then consider two different scenarios. One in which we gauge $SU(M)_L$ and one in which we only gauge a subgroup $SU(2)_L \subset SU(M)_L$. 

We discover a rich phase diagram that contains one, two or three fixed points and determine the associated renormalisation group flow. A phase featuring complete asymptotic freedom and simultaneously a complete (in both couplings) interacting infrared fixed point emerges. Intriguingly we further expose the emergence of special renormalisation group trajectories along which one coupling displays asymptotic freedom and the other asymptotic safety, while both flow in the infrared to an interacting fixed point. We term these trajectories  {\it safety free}. 

We then present the actual running of the couplings for potentially phenomenologically interesting renormalisation group trajectories that hint at new avenues for model building.  Among these also new ways to generate near-conformal dynamics a l\'a walking.

The paper is organized as follows. In Section \ref{sec:NM} we study the phase diagram for the $SU(N) \times SU(M)_L$ gauge theory while in Section \ref{sec:N2} we study the theory with $SU(2)_L \subset SU(M)_L$ gauged. We sketch-out phenomenological implications of the quantum critical behaviour of the theories investigated here in Section \ref{pheno}. Finally we present our conclusions in Section \ref{conclusions}.

\section{The Quantum Critical Behaviour of the $SU(N)\times SU(M)_L$ Theory}\label{sec:NM}

We start out by considering the simple gauge theory with an $SU(N)$ gauge group and two sets of $M$ Weyl fermions $q$ and $\tilde{q}$ in the fundamental and antifundamental representations respectively. This theory has an $SU(M)_L\times SU(M)_R \times U(1)_V$ anomaly free global symmetry. At the classical level there is an additional abelian $U(1)$ symmetry but this is broken by an anomaly at the quantum level. This theory is QCD with $N$ colors and $M$ Dirac flavors. This theory is asymptotically free if $M<\frac{11}{2}N$ and possesses an infrared fixed point for a number of flavours just below this critical value.

We now gauge the $SU(M)_L$ part of the global flavor symmetry so that the theory has the semi-simple gauge group $SU(N)\times SU(M)_L$. The Weyl fermions $q$ are charged under $SU(N)\times SU(M)_L$ while the fermions $\tilde{q}$ are charged only under $SU(N)$. For $M>2$ this therefore induces an $SU(M)_L^3$ gauge anomaly which we need to avoid. To cancel the anomaly we add $N$ Weyl fermions $L$ in the antifundamental representation of $SU(M)_L$ since the fermions $q$ belong to the fundamental representation of $SU(M)_L$. There are no other gauge anomalies. Any other mixed anomaly must contain a single $SU(N)$ or $SU(M)_L$ gauge current and must vanish since it is proportional to the trace of the associated symmetry generator. 

In the special case where $M=2$, which is closer to the Standard Model, one can actually gauge the $SU(2)_L$ part directly without inducing any gauge anomalies, provided $N$ is even to avoid Witten's topological anomaly \cite{Witten:1982fp}. Hence there is no need to add additional matter in this case. As mentioned above for the generic $M>2$ case we add the additional set of fermions ($L$ for leptons) to avoid anomalies. In Table  \ref{tab:1} we summarise the matter content of the theory.
\begin{table}[htbp]
\center
\begin{tabular}{l| c c c c c c}
 & $\left[SU(N)\right]$ & $\left[SU(M)_L\right]$ & $SU(M)_R$&  $SU(N)$ & $U(1)_V$ \\ \hline
  $q$ & $\square$ & $\square$ & 1 & 1 & +1\\
 $\tilde{q}$ & $\overline{\square}$ & 1 & $\overline{\square}$ &  1 & -1 \\
 $L$ & 1 & $\overline{\square}$ & 1 & $ \overline \square$ & -1 
\end{tabular}
\caption{Matter content and symmetries}\label{tab:1}
\end{table}
The brackets denote the gauge symmetries while $SU(M)_R \times SU(N)$ is the remaining nonabelian global symmetry. The theory enjoys only a single anomaly free abelian $U(1)_V$ global symmetry despite the fact that at the classical level it possesses three. However since the gauge group is semi-simple the global symmetry current can couple to both $SU(N)$ and $SU(M)_L$ gauge currents respectively. The anomalies have the potential to arise via the following diagrams
\vspace{0.05\linewidth}
\begin{figure}[htbp]
\begin{minipage}[b]{0.35\linewidth}
\centering
\begin{fmffile}{triangleN}
		\begin{fmfgraph*}(100,50)
			\fmfleft{i1}
			\fmfright{r1,r2}
			\fmftop{o1}
			\fmfbottom{o2}
			\fmf{fermion}{i1,o1}
			\fmf{fermion}{o2,i1}
			\fmf{fermion}{o1,o2}
			\fmf{boson}{o1,r2}
			\fmf{boson}{o2,r1}
			\fmfdot{i1}
			\fmflabel{$SU(N)$}{r1}
			\fmflabel{$SU(N)$}{r2}
		\end{fmfgraph*}
\end{fmffile}
\end{minipage}
\begin{minipage}[b]{0.35\linewidth}
\centering
\begin{fmffile}{triangleM}
		\begin{fmfgraph*}(100,50)
			\fmfleft{i1}
			\fmfright{r1,r2}
			\fmftop{o1}
			\fmfbottom{o2}
			\fmf{fermion}{i1,o1}
			\fmf{fermion}{o2,i1}
			\fmf{fermion}{o1,o2}
			\fmf{boson}{o1,r2}
			\fmf{boson}{o2,r1}
			\fmfdot{i1}
			\fmflabel{$SU(M)$}{r1}
			\fmflabel{$SU(M)$}{r2}
		\end{fmfgraph*}
\end{fmffile}
\end{minipage}
\end{figure}
\vspace{0.05\linewidth}

These diagrams, however, vanish provided that the fermions are charged as in the above Table \ref{tab:1}. There can be no other abelian anomaly free symmetries. If any one fermion is charged under an abelian symmetry then all three must be charged due to their gauge symmetry assignments. This can only be the one already given in the Table. Lastly we provide the Lagrangian of the system
\beq
\begin{split}
\mathcal L  = & -\frac{1}{4}G^a_{\mu\nu}G^{a\mu\nu} -\frac{1}{4}F^i_{\mu\nu}F^{i\mu\nu}\\
& +\overline q^{n_c m_c} \overline \sigma^\mu \left( \delta^{n_c'}_{n_c} \delta^{m_c'}_{m_c} \partial_\mu - i g_N (T^a)^{n_c'}_{n_c}  \delta^{m_c'}_{m_c} G^a_\mu - i g_M (S^i)^{m_c'}_{m_c} \delta^{n_c'}_{n_c} A^i_\mu \right) q_{n_c' m_c'}\\
& + \overline{\tilde q}_{n_c m_f} \overline \sigma^\mu \left(  \delta^{n_c}_{n_c'} \delta^{m_f}_{m_f'} \partial_\mu + i g_N (T^a)^{n_c}_{n_c'}  \delta^{m_f}_{m_f'} G^a_\mu \right) \tilde{q}^{n_c' m_f'}\\
& + \overline{L}_{n_f m_c} \overline \sigma^\mu \left( \delta_{n_f'}^{n_f} \delta_{m_c'}^{m_c} \partial_\mu + i g_M  (S^i)^{n_f}_{n_f'} \delta_{m_c'}^{m_c} A^i_\mu \right) L^{n_f' m_c'},\\
\end{split}
\eeq
where $n_c, m_c$ refer to the "color" indices of the gauge groups $SU(N), SU(M)_L$ and $n_f, m_f$ refer to the flavor indices of the global symmetry groups $SU (N), SU(M)_R$. The gauge field strengths, gauge fields and generators of $SU(N)$ and $SU(M)_L$ are denoted as $G_{\mu\nu}^a,\ G_{\mu}^a,\ T^a$ and $F_{\mu\nu}^i,\ A_{\mu}^i,\ S^i$ respectively. The two gauge couplings are denoted as $g_N$ and $g_M$.


The calculation of the beta functions up to second loop-order for a semi-simple gauge theory was first calculated in \cite{Jones:1981we}. It was further generalized to semi-simple gauge-Yukawa theories in \cite{Luo:2002ti}. In our case with two gauge couplings they are
\begin{eqnarray}
\beta_N\left(\alpha_N,\alpha_M\right)&=&-a_N\frac{\alpha^2_N}{2\pi}-b_N\frac{\alpha^3_N}{(2\pi)^2}-c_N\frac{\alpha^2_N\alpha_M}{(2\pi)^2},\\
\beta_M\left(\alpha_M,\alpha_N\right)&=&-a_M\frac{\alpha^2_M}{2\pi}-b_M\frac{\alpha^3_M}{(2\pi)^2}-c_M\frac{\alpha^2_M\alpha_N}{(2\pi)^2}.
\end{eqnarray}
where $\alpha_{N} = \frac{g_{N}^2}{4\pi}$ and $\alpha_{M} = \frac{g_{M}^2}{4\pi}$. The beta function coefficients for this theory can be found in Appendix \ref{app:NM}. Asymptotic freedom (AF) is dictated by the sign of the first coefficients $a_N$ and $a_M$. In principle there are four possibilities 
\begin{eqnarray}
&& a_N >0 \ , \qquad a_M>0 \ , \qquad \text{both couplings are AF} \\
&& a_N<0 \ , \qquad a_M>0 \ , \qquad \text{only $\alpha_M$ is AF} \\
&& a_N>0 \ , \qquad a_M<0 \ , \qquad \text{only $\alpha_N$ is AF} \\
&& a_N<0 \ , \qquad a_M<0 \ , \qquad \text{None are AF}
\end{eqnarray}
However the last situation where both gauge interactions are infrared free cannot be realised since the two conditions $a_N<0$ and $a_M<0$ imply $\frac{2}{11}>\frac{M}{N}$ and $\frac{M}{N}>\frac{11}{2}$ which cannot be simultaneously satisfied. Hence we are left with three regions in which either only one of the gauge interactions is asymptotically free or both are asymptotically free. 

\begin{figure}[htbp]
\center
\includegraphics[width=0.6\textwidth]{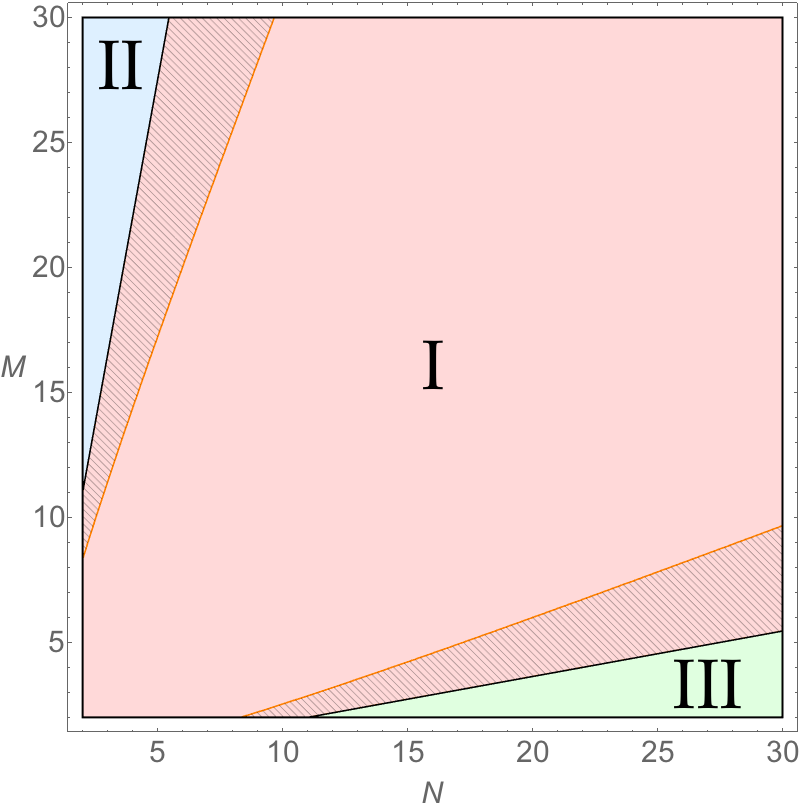}
\caption{The three regions where I) both couplings $\alpha_N$ and $\alpha_M$ are asymptotically free (red), II) where $\alpha_M$ ($\alpha_N$) is asymptotically (infrared) free (blue), III) where $\alpha_N$ ($\alpha_M$) is asymptotically (infrared) free (green). The hatched regions indicate where FP$_1$ (the upper region) and FP$_2$ (the lower region) are physical.}\label{NMAF}
\end{figure}

We plot these three regions in the $(N,M)$ plane in Fig. \ref{NMAF}. Region I (red) is bounded by $\frac{2}{11}<\frac{M}{N}<\frac{11}{2}$ while region II (blue) is bounded by $\frac{11}{2}<\frac{M}{N}$ and region III (green) is bounded by $\frac{M}{N}<\frac{2}{11}$. We will now discuss fixed points in each region.

Setting to zero the two coupled beta functions we find that there are nontrivial fixed points located at
\begin{eqnarray}
&\quad \alpha_{N}^* = -2\pi \frac{a_N}{b_N}, \quad   \alpha_{M}^* = 0 &  \qquad (\text{FP}_1)  \\
&\alpha_{N}^*= 0 \ , \quad \alpha_{M}^* = -2\pi \frac{a_M}{b_M} &   \qquad (\text{FP}_2) \\
&\alpha_{N}^* = - 2\pi \frac{a_Nb_M - a_Mc_N}{b_Nb_M - c_Nc_M} \ , \quad \alpha_{M}^* = -2\pi \frac{a_Mb_N - a_N c_M}{b_Nb_M - c_Nc_M}& \qquad (\text{FP}_3) 
\end{eqnarray}
In Appendix \ref{app:fixed} we show that the third fixed point does not exist in neither region I, II or III for this specific theory. Hence we are only left with the first and second nontrivial fixed points. If we switch off the gauge coupling $\alpha_N$ ($\alpha_M$) then FP$_2$ (FP$_1$) is  the infrared Banks-Zaks fixed point for $\alpha_M$ ($\alpha_N$) provided that $\alpha_M$ ($\alpha_N$) is asymptotically free. This, of course, also implies that they cannot be ultraviolet fixed points in region II and III respectively and falls in line with the results of \cite{Litim:2014uca} where an ultraviolet fixed point is generated perturbatively if the theory contains scalars.

The phase diagram of this specific theory is therefore rather simple. There are no further nontrivial fixed points induced by the mixing of the gauge couplings. On the other hand we will show below that a theory where this is the case indeed does exists. In Fig. \ref{NMAF} we plot the phase diagram in the $(N,M)$ plane. In order for the fixed points to be physically acceptable we demand perturbation theory to hold and therefore require $\alpha_N^*<1$ and $\alpha_M^*<1$. This is the hatched region in Fig. \ref{NMAF}. 

\begin{figure}[htbp]
\begin{minipage}[b]{0.45\linewidth}
\centering
\includegraphics[width=\textwidth]{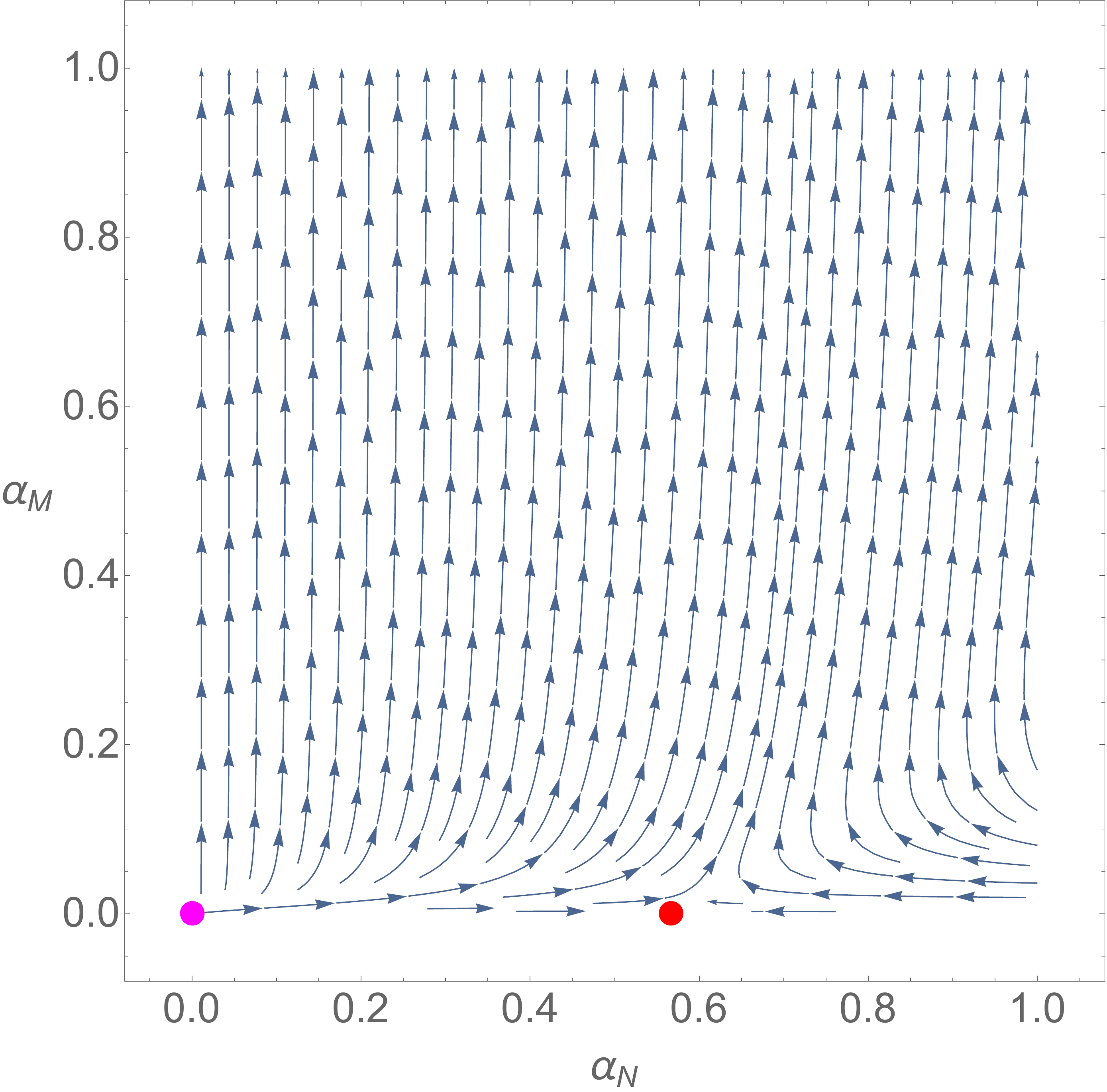}
\end{minipage}
\hspace{0.5cm}
\begin{minipage}[b]{0.45\linewidth}
\centering
\includegraphics[width=\textwidth]{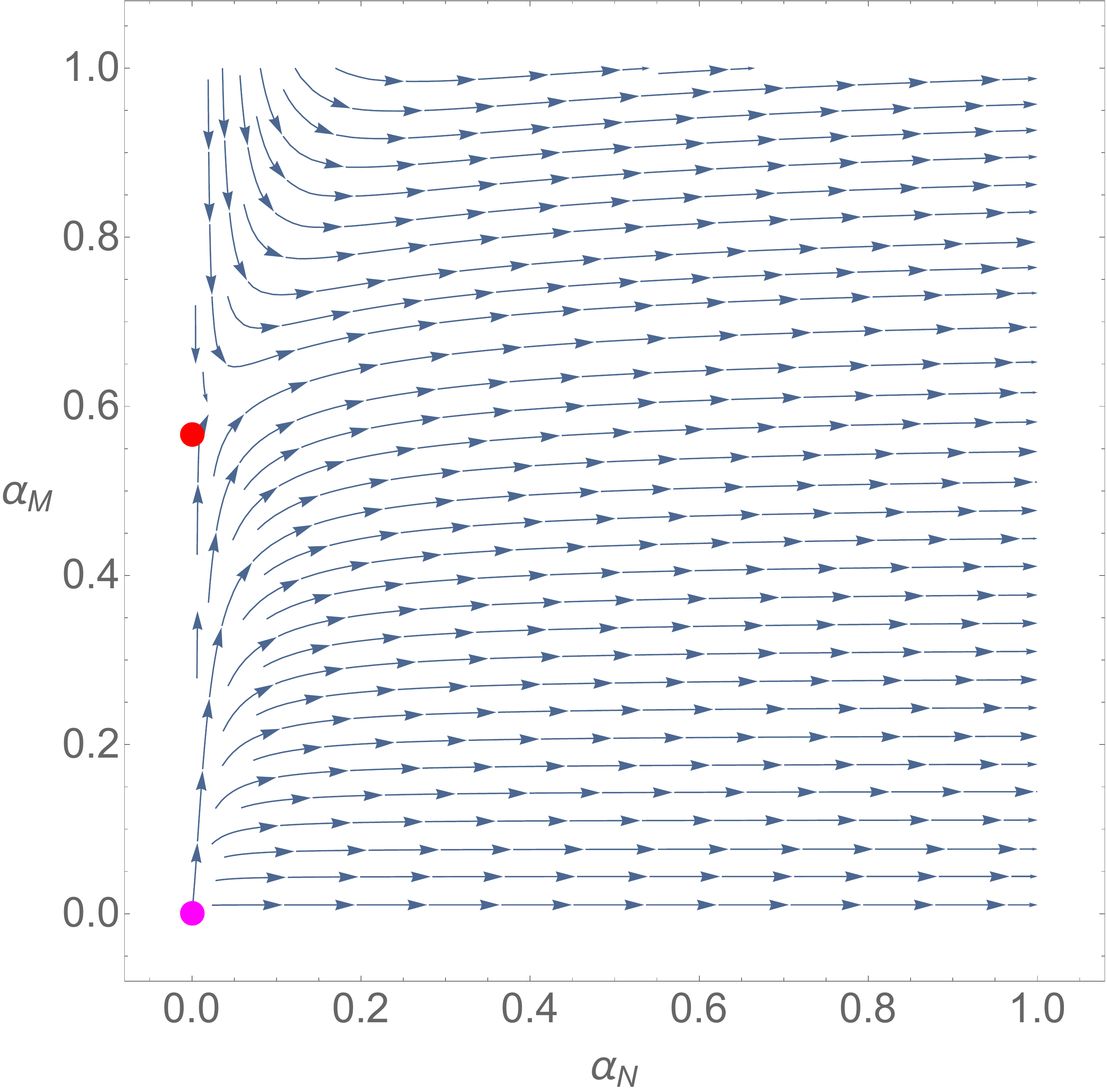}
\end{minipage}
\caption{The left (right) panel shows the phase diagram of the theory with $N=7$ and $M=25$ ($N=25$ and $M=7$) in the region where only FP$_1$ (FP$_2$) exists. The violet point is the ultraviolet free fixed point while the red point is FP$_1$ (left) or FP$_2$ (right). }\label{FPflow}
\end{figure}

We now discuss the stability of the two fixed points FP$_1$ and FP$_2$ by linearising the flow around the fixed points and determining the eigenvalues of
\begin{eqnarray}
M =
\left( 
\begin{array}{cc}
\frac{\partial \beta_N}{\partial \alpha_N} & \frac{\partial \beta_N}{\partial \alpha_M}  \\
\frac{\partial \beta_M}{\partial \alpha_N} & \frac{\partial \beta_M}{\partial \alpha_M} 
\end{array}
\right)_{|\alpha_N=\alpha_N^*,\ \alpha_M=\alpha_M^*}.\label{eq:stabm}
\end{eqnarray}
If an eigenvalue is negative the fixed point is unstable (relevant direction)  while if it is positive it is stable (along their associated eigendirections) and leads to an irrelevant direction.  At the two fixed points FP$_1$ and $FP_2$ they are
\begin{eqnarray}
\text{Eigenvalues}(M_{\text{FP}_1}) &=& \left(\frac{2(2M-11N)^2N}{3(13MN^2 - 34N^3-3M)} ,0  \right) \\
\text{Eigenvalues}(M_{\text{FP}_2}) &=& \left(0, \frac{2(11M-2N)^2M}{3(34M^3 +3N -13M^2 N)} \right)
\end{eqnarray}
The nonzero eigenvalue of $M_{\text{FP}_1}$ is positive for $\alpha_N^*<1$ and the nonzero eigenvalue of $M_{\text{FP}_2}$ is positive for $\alpha_M^*<1$. Hence if the fixed points exist then they are attractive along one direction. The stability along the remaining eigendirection cannot be determined. The vanishing of one of the eigenvalues is easily understandable because it is the footprint of the gaussian behaviour with respect to that coupling direction. In Fig. \ref{FPflow} we plot the flow of the couplings for two sets of generic values of $N$ and $M$. In both plots the trivial ultraviolet fixed point is marked violet. In the left plot the red point is FP$_1$ while in the right plot the red point is FP$_2$. We see that for all the asymptotically free trajectories the couplings will grow large as the infrared regime is approached. The only exception in the left (right) plot is the trajectory for which $\alpha_M$ ($\alpha_N$) is switched off. Here $\alpha_N$ ($\alpha_M$) just approaches the fixed point in the infrared. If the system evolves along one of the trajectories that come close to either of the two nontrivial fixed points the couplings will exhibit near scale invariant characteristics (i.e. walking) at intermediate  scales before blowing up in the deep infrared. 

Finally we observe that there is a rather special trajectory which flows directly out of FP$_1$ in the left plot and directly out of FP$_2$ in the right plot. In the left plot along this trajectory the fixed point FP$_1$ acts as a trivial ultraviolet fixed point for $\alpha_M$ but as a nontrivial ultraviolet fixed point for $\alpha_N$. Hence the coupling $\alpha_M$ is asymptotically free but $\alpha_N$ is asymptotically safe. Such a trajectory is a \emph{safety free} trajectory. The same occurs in the right plot but with the role of $\alpha_N$ and $\alpha_M$ switched.

\section{Gauging an $SU(2)_L$ subgroup enriches the Phase Diagram}\label{sec:N2}

We will now gauge only an $SU(2)_L$ subgroup of the $SU(M)_L$ group and remember that the original theory was composed of an $SU(N)$ gauge group with $M$ fundamental Weyl fermions $q$ and $M$ antifundamental fermions $\tilde{q}$. The original theory has an $SU(M)_L \times SU(M)_R \times U(1)_V$ anomaly free global symmetry. If we only gauge an $SU(2)_L$ subgroup of $SU(M)_L$ we do not run into the problem of inducing gauge anomalies as compared to the above case where we gauged the entire $SU(M)_L$ since all representations are now (pseudo)real. Therefore from the point of view of gauge anomalies we do not need to add additional fermions to the theory. 

In the Standard Model the gauge anomalies vanish only due to a nontrivial cancelation between the quark and lepton contributions. However in the Standard Model also an $U(1)_Y\subset SU(M)_R$ abelian subgroup is gauged and this induces mixed anomalies that can only be canceled by the inclusion of leptons. This is not an issue here.

However since one of the gauge groups is now $SU(2)$ we have to worry about the Witten topological anomaly \cite{Witten:1982fp}. So far the theory contains $N$ Weyl doublets and hence only in the specific case where $N$ is even does the Witten anomaly vanish. In general however we can cancel the Witten anomaly by including $N$ lepton doublets such that the theory contains $2N$ doublets for any $N$. In Table \ref{tab:2} we summarise the matter content of the theory. Note that there are three abelian anomaly free symmetries.

\begin{table}[htbp]
\center
\begin{tabular}{c|cccccccc}
 & $[SU(N)]$ & [$SU(2)_L$] & $SU(M-2)_L$ & $SU(M)_R$ & $SU(N)$ &  $U(1)_1$ & $U(1)_2$ & $U(1)_3$ \\
 \hline
 $q$ & $\square$ & $\square$  & 1 & 1 & 1 & $M$  & $2-M$ & 0  \\
 $q'$ & $\square$ & 1 &  $\square$ & 1 & 1 &  $0$ & $2$ & -M \\
 $\tilde{q}$  & $\overline{\square}$ & 1 &1  & $\overline{\square}$ & 1 & $-2$  & 0 & M-2 \\
 $L$ & 1 & $\square$ & 1 & 1 & $\overline{\square}$ & $-M$ & $M-2$ & 0 
  \end{tabular}
\caption{Matter content of the theory with an $SU(2)_L$ gauged subgroup. }\label{tab:2}
\end{table}

The evolution of the system is characterised to two loops by the coupled beta functions 
\begin{eqnarray}
\beta_N &=& - a_N \frac{\alpha_N^2}{2\pi} -b_N \frac{\alpha_N^3}{(2\pi)^2} - c_N \frac{\alpha_2 \alpha_N^2}{(2\pi)^2} \\
\beta_2 &=& - a_2 \frac{\alpha_2^2}{2\pi} -b_2 \frac{\alpha_2^3}{(2\pi)^2} - c_2 \frac{\alpha_N \alpha_2^2}{(2\pi)^2}
\end{eqnarray}
where the beta function coefficients can be found  in Appendix \ref{app:N2L}. Asymptotic freedom is dictated by the sign of the first coefficients $a_N$ and $a_2$. This gives four different possibilities 
\begin{eqnarray}
&& a_N >0 \ , \qquad a_2>0 \ , \qquad \text{both couplings are AF} \\
&& a_N<0 \ , \qquad a_2>0 \ , \qquad \text{only $\alpha_2$ is AF} \\
&& a_N>0 \ , \qquad a_2<0 \ , \qquad \text{only $\alpha_N$ is AF} \\
&& a_N<0 \ , \qquad a_2<0 \ , \qquad \text{None are AF}
\end{eqnarray} 
These four conditions map out four distinct regions in the $(N,M)$ plane. We now proceed to study the fixed point structure of the theory. These are
\begin{eqnarray}
&\quad \alpha_{N}^* = -2\pi \frac{a_N}{b_N}, \quad   \alpha_{2}^* = 0 &  \qquad (\text{FP}_1)  \\
&\alpha_{N}^*= 0 \ , \quad \alpha_{2}^* = -2\pi \frac{a_2}{b_2} &   \qquad (\text{FP}_2) \\
&\alpha_{N}^* = - 2\pi \frac{a_Nb_2 - a_2c_N}{b_Nb_2 - c_Nc_2} \ , \quad \alpha_{2}^* = -2\pi \frac{a_2b_N - a_N c_2}{b_Nb_2 - c_Nc_2}& \qquad (\text{FP}_3) 
\end{eqnarray}
Besides the trivial fixed point there are three nontrivial fixed points denoted by FP$_1$, FP$_2$ and FP$_3$. In terms of $N$ and $M$ they are
\begin{eqnarray}
& \alpha_N^* =  -2\pi \frac{2(2MN-11N^2)}{13MN^2-34N^3-3M}   \ , \qquad \alpha_2^*=0 & \qquad \qquad (\text{FP}_1) \\
& \alpha_N^* =  0   \ , \qquad \alpha_2^*= - 2 \pi \frac{8(N-11)}{49N-272} & \qquad \qquad (\text{FP}_2) \\
& \alpha_N^* =  -2\pi \frac{4(98MN^2+2974N^2+198N-539N^3-544MN)}{1274MN^3+ 18469N^3+1632M +27N-3332N^4-7072MN^2-294MN}   \ ,  & \qquad \qquad (\text{FP}_3) \\
&\alpha_2^*=  - 2 \pi \frac{4(46MN^3+1496N^3+132M-103N^4-572MN^2- 33 N^2-6MN)}{1274MN^3+ 18469N^3+1632M +27N-3332N^4-7072MN^2-294MN}  & \nonumber
\end{eqnarray}
We discover  four different regions bounded by
\begin{eqnarray}
&\text{Region I}, \quad N<11, \   M< \frac{11}{2}N\;,& \\
&\text{Region II}, \quad N< 11 , \   M>\frac{11}{2}N\;,& \\ 
&\text{Region III}, \quad N>11 , \  M<\frac{11}{2}N\;,& \\ 
&\text{Region IV}, \quad N>11 , \  M>\frac{11}{2}N\;,&
\end{eqnarray}

We will require that the values of the fixed points are larger than zero and less than unity $0< \alpha_N^*<1$ and $0< \alpha_2^*<1$. These requirements will be enforced on all three fixed point solutions FP$_1$, FP$_2$ and FP$_3$. In our analysis each fixed point can only be trusted if it is perturbative.

\begin{figure}
\includegraphics[width=0.6\textwidth]{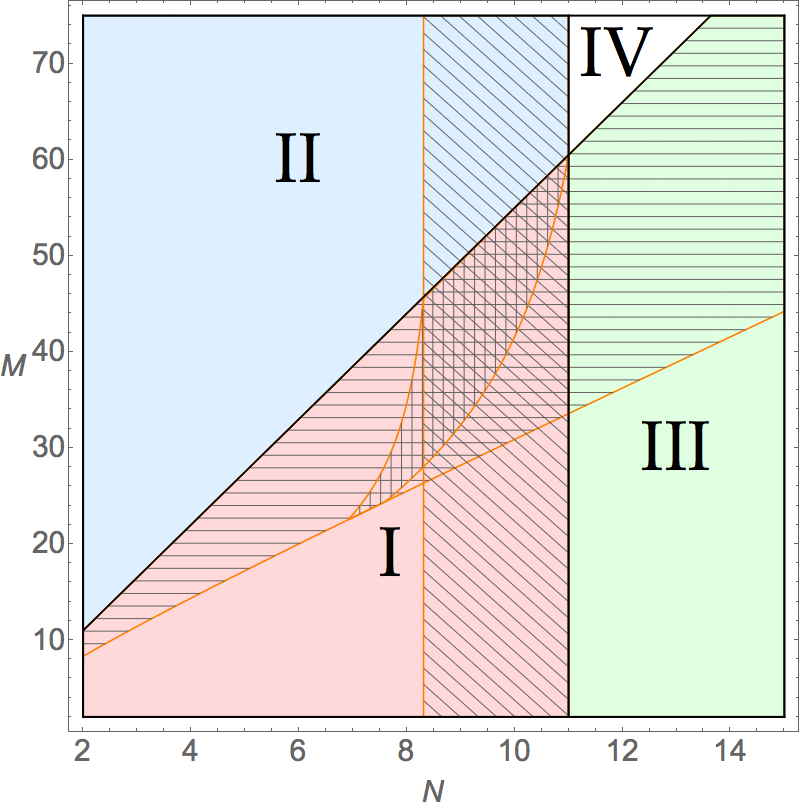}
\caption{The four regions where in region I (IV) both couplings are asymptotically (infrared) free and in region II (III) $\alpha_N$ ($\alpha_2$) is infrared free while $\alpha_2$ ($\alpha_N$) is asymptotically free. The hatched regions mark where the fixed points FP$_1$ (horizontal lines), FP$_2$ (diagonal lines) and FP$_3$ (vertical lines) exists. }
\label{fig:n2}
\end{figure}

In Fig. \ref{fig:n2} we plot all four regions together with the regions in which there exists either one, two or three fixed points. The black solid lines correspond to $N=11$ and $M=\tfrac{11}{2}N$ and separate the four different regions. The orange lines separate the regions in which one, two or all three fixed points exist from the regions in which none exists. If some or all fixed points are physical the corresponding region is hatched. The theories that lie in a region which is either single, double or triple hatched possess either one, two or all three fixed points respectively. Finally the regions in which FP$_1$, FP$_2$ or FP$_3$ exist are marked with horizontal, diagonal or vertical lines respectively. 

One should note that if the third nontrivial fixed point FP$_3$  exists then so does either FP$_1$ and FP$_2$ or only FP$_1$. It never exists simultaneously with only FP$_2$. The theories that can settle at FP$_3$ in the deep infrared are the ones lying in the upper right part of Region I marked by the black diagonal line and the two orange curved lines in Fig. \ref{fig:n2}. 

The fixed points can be either stable, unstable or metastable. Similar to above we classify the fixed points according to their stability by linearising the beta functions around the fixed points and study the eigenvalues of the matrix in  \eqref{eq:stabm} where $\beta_M$ and $\alpha_M$ is now $\beta_2$ and $\alpha_2$.

Consider the first two fixed points FP$_1$ and FP$_2$. Here the stability matrix always has a zero as one of its eigenvalues. The other eigenvalue is always positive at both fixed points with eigenvector $(1,0)^T$ at FP$_1$ and eigenvector $(0,1)^T$ at FP$_2$. Therefore FP$_1$ is always attractive in the direction of $\alpha_N$ while FP$_2$ is always attractive in the direction of $\alpha_2$. 

Consider now instead the third fixed point FP$_3$. This is the nontrivial fixed point which emerges due to the mixing between the couplings in the beta functions. In the region where the fixed point value of the couplings are positive and less than unity both eigenvalues of the stability matrix are positive implying that the fixed point is infrared attractive from all directions. 

There are special renormalisation group lines connecting FP$_3$ to either FP$_1$ or FP$_2$. Along these lines one of the two couplings moves from an infrared fixed point to a gaussian (asymptotically free) fixed point while the other reaches an interacting ultraviolet  (asymptotically safe) fixed point. These are lines of safety free theories.  This new phase is an addition to the completely asymptotically free or safe case studied earlier in the literature.

\begin{figure}[h]
\begin{minipage}{0.45\textwidth}
\centering
\includegraphics[width=0.9\textwidth]{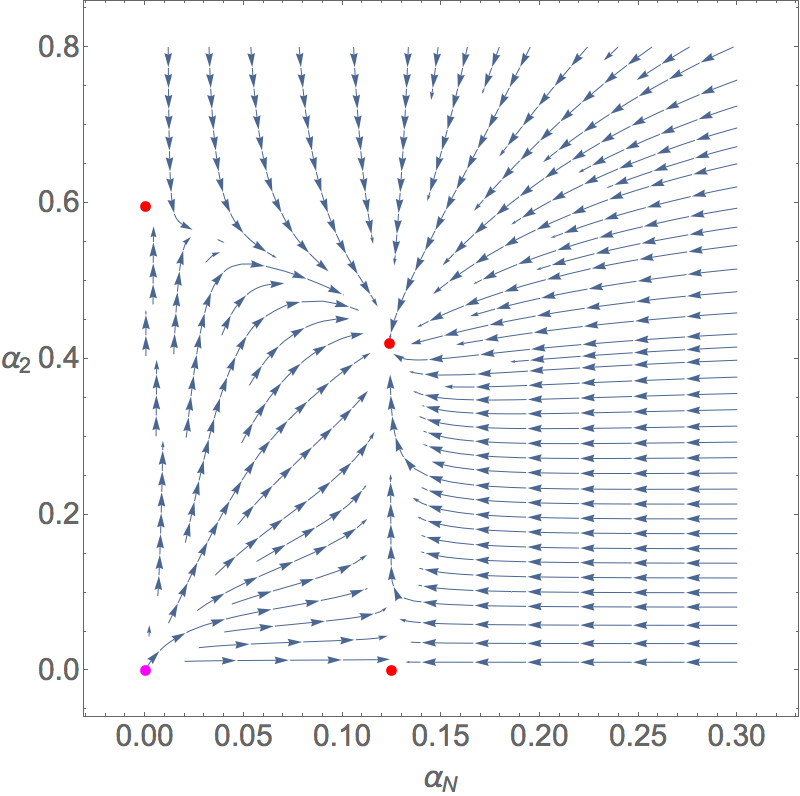}
\end{minipage}
\begin{minipage}{0.45\textwidth}
\centering
\includegraphics[width=0.9\textwidth]{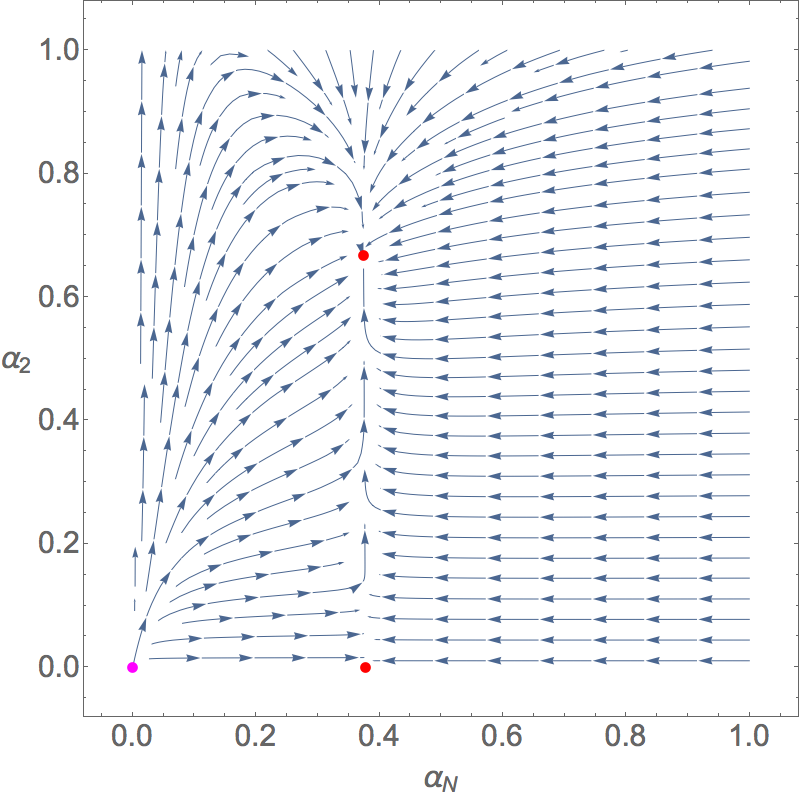}
\end{minipage}
\caption{Coupling flows of theories in region I: The left plot shows the $N=9$ and $M=40$ theory with fixed points FP$_1$, FP$_2$ and FP$_3$. The right plot shows the $N=8$ and $M=30$ theory with fixed points FP$_1$ and FP$_3$. }
\label{fig:flow13123}
\end{figure}

It is illustrative to plot the flow of the couplings for a variety of different theories having either one, two or all three fixed points. As a first example we plot the flow for $N=9$ and $M=40$ in the left plot of Fig. \ref{fig:flow13123}. As can be seen the two fixed points FP$_1$ and FP$_2$ are both attractive along one direction. This is the eigendirection with positive eigenvalue of the stability matrix which is in the $\alpha_N$ direction for FP$_1$ and in the $\alpha_2$ direction for FP$_2$. Along the other eigendirection with vanishing eigenvalue both fixed points are repulsive for positive values of the couplings. The eigenvalues at these two fixed points are,
\begin{eqnarray}
\text{Eigenvalues}(M_{\text{FP}_1}) &=& \left(\frac{2(2M-11N)^2N}{3(13MN^2 - 34N^3-3M)} ,0  \right), \\
\text{Eigenvalues}(M_{\text{FP}_2}) &=& \left(0, \frac{16(N-11)^2}{3(49N-272)} \right).
\end{eqnarray}

 There is also the nontrivial fixed point FP$_3$ which is infrared attractive in all directions. This is the fixed point to which the theory will finally settle in the deep infrared. The eigenvalues at this fixed point is given below but due to the complicated nature of this eigenvalue in terms of $N$ and $M$ we only show the numerical values with $N=9$ and $M=40$,
\beq
\text{Eigenvalues}(M_{\text{FP}_3})= \left( 0.0622, 0.1243 \right).
\eeq
The second example is the flow of the theory with $N=8$ and $M=30$. This is the right plot of Fig. \ref{fig:flow13123} and corresponds to the region with double hatched vertical and horizontal lines of Fig. \ref{fig:n2}. In this theory the fixed points FP$_1$ and FP$_3$ exist simultaneously. The flow of the theory for these two fixed points is similar to the flow of the fixed points in the example above i.e. FP$_1$ is attractive along one direction and repulsive in the other direction while FP$_3$ is attractive along all directions. It is important to note that the fixed point FP$_2$ exists along the direction of $\alpha_2$. However we will disregard this fixed point since here the value of the coupling is larger than one. Thus along the trajectories for which the couplings stay below unity the theory will settle at FP$_3$ in the deep infrared.
 
\begin{figure}
\begin{minipage}{0.30\textwidth}
\centering
\includegraphics[width=0.9\textwidth]{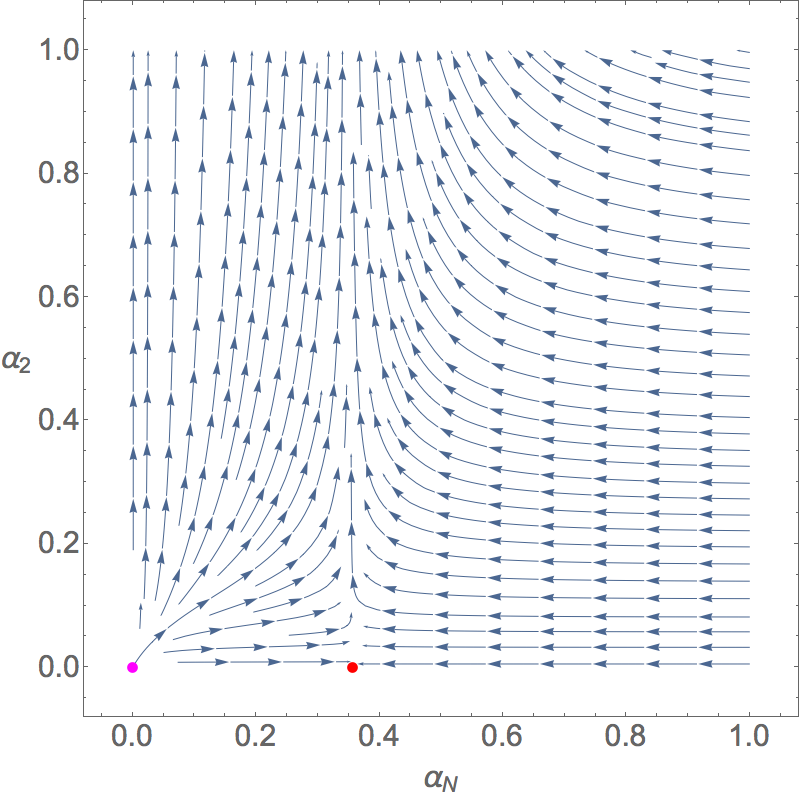}
\end{minipage}
\begin{minipage}{0.30\textwidth}
\centering
\includegraphics[width=0.9\textwidth]{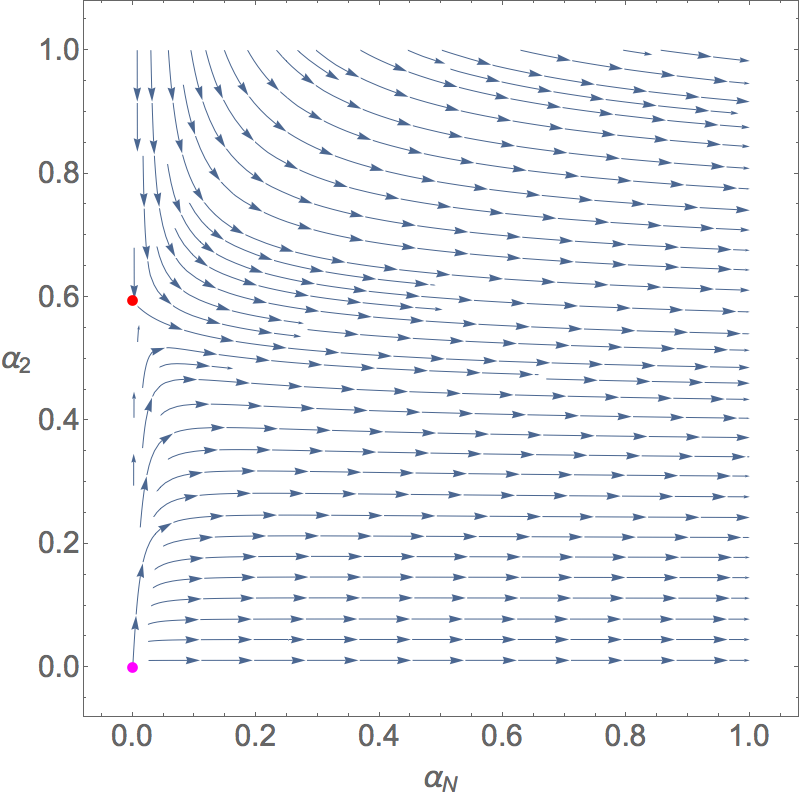}
\end{minipage}
\begin{minipage}{0.30\textwidth}
\centering
\includegraphics[width=0.9\textwidth]{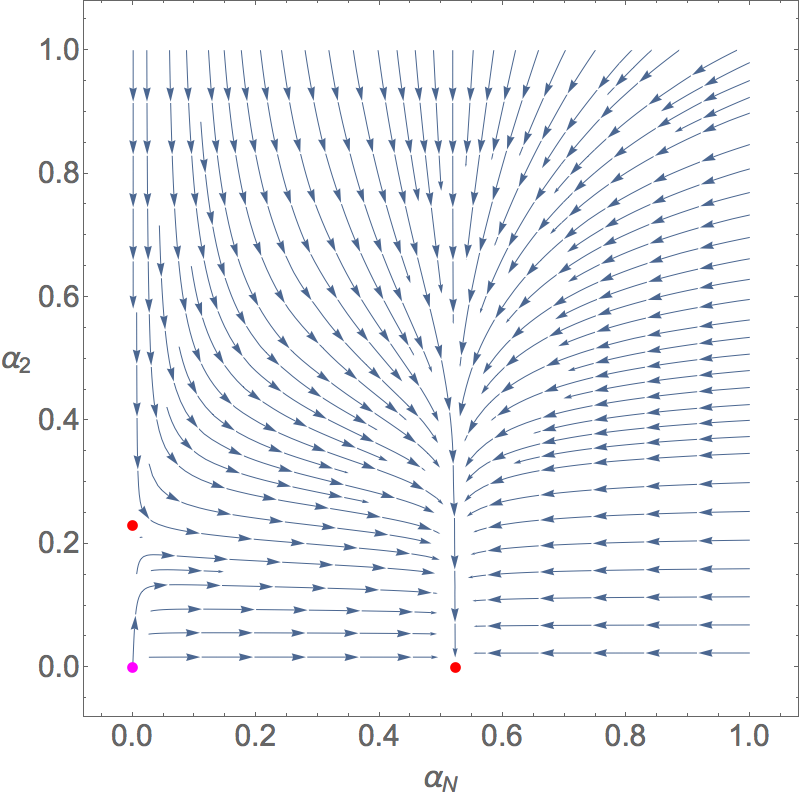}
\end{minipage}
\caption{Coupling flows of theories in region I: The left plot shows the $N=6$ $M=24$ theory with fixed point FP$_1$. The center plot shows the $N=9$ $M=18$ theory with fixed point FP$_2$. The right plot shows the $N=10$ and $M=34$ theory with fixed points FP$_1$ and FP$_2$.}
\label{fig:flow1212}
\end{figure}

The third and fourth example are of theories where only FP$_1$ or FP$_2$ exists. This is the single hatched parts of region I. In Fig \ref{fig:flow1212} the left plot corresponds to a $N=6$ and $M=24$ theory with fixed point FP$_1$ and the center plot is a $N=9$ and $M=18$ theory with fixed point FP$_2$. In both of these examples the fixed point FP$_1$ (FP$_2$) is attractive along one direction $\alpha_N$ ($\alpha_2$) and repulsive along the other.

The fifth example is of a $N=10$ and $M=34$ theory where both fixed points FP$_1$ and FP$_2$ exists simultaneously. This is the double hatched part of region I with horizontal and diagonal lines. The fixed point FP$_1$ (FP$_2$) is attractive along the associated eigendirection of $\alpha_N$ ($\alpha_2$) while being repulsive along the other. Furthermore the fixed point FP$_1$ is attractive along directions of postive $\alpha_2$ but repulsive in the direction of negative $\alpha_2$. This is of course highly unstable and as the theory eventually flows to negative values of $\alpha_2$ it is ill defined.

Region II is bounded by $N<11$ and $M>\frac{11}{2}N$. Similar to the analysis of region I we shall demand that the fixed points are positive and less than unity in order to trust perturbation theory. In region II the coupling $\alpha_2$ is asymptotically free while $\alpha_N$ is not. If the first fixed point exists then it must be a UV fixed point. However since the value of $\alpha_N^*$ is negative this does not occur. There is a region in which the second fixed point FP$_2$ exists. At this point one of the eigenvalues of the stability matrix vanishes while the other is nonzero. Lastly we remark that the third fixed point FP$_3$ is not positive and hence does exist anywhere in region II.   

\begin{figure}[ht]
\begin{minipage}[b]{0.45\linewidth}
\centering
\includegraphics[width=\textwidth]{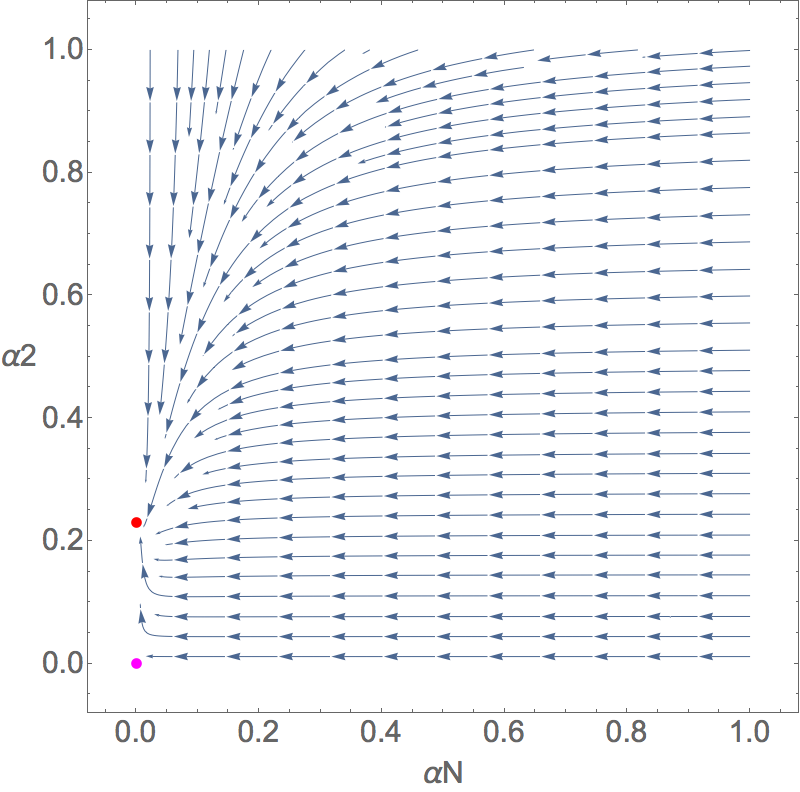}
\end{minipage}
\hspace{0.5cm}
\begin{minipage}[b]{0.45\linewidth}
\centering
\includegraphics[width=\textwidth]{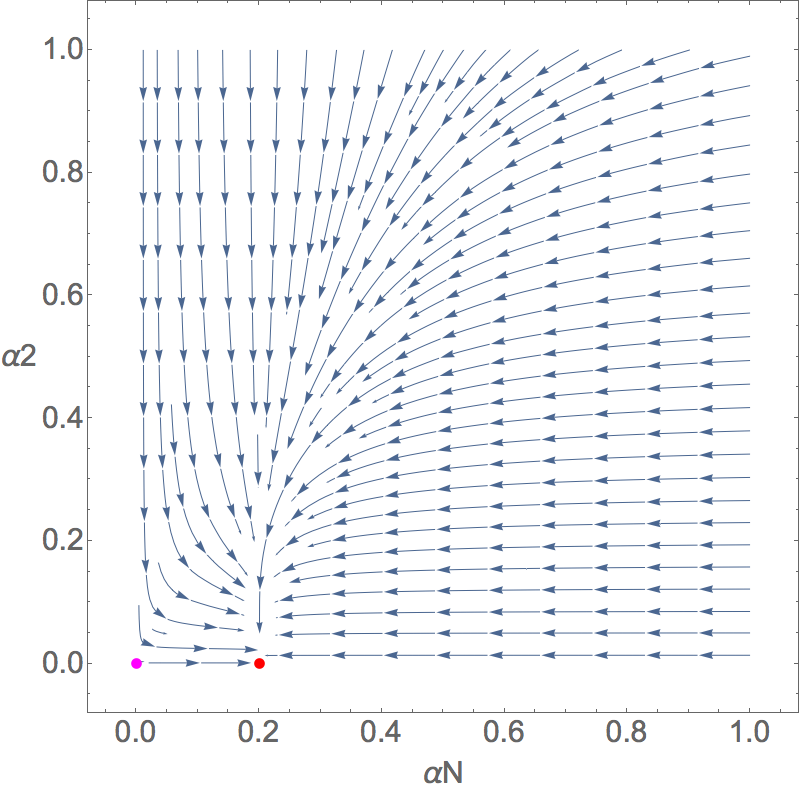}
\end{minipage}
\caption{Coupling flows of theories in region II and III: The left plot shows the $N=10$ and $M=60$ theory with fixed point FP$_2$. The right plot shows the $N=13$ and $M=50$ theory with fixed point FP$_2$. }
\label{Flow}
\end{figure}

In the left plot of Fig. \ref{Flow} we show the flow of the couplings for $N=10$ and $M=60$. In the far UV the theory sits in the lower right corner of the flow. As the theory evolves towards the IR it develops a perturbative IR fixed point. As can be seen the fixed point happens to be attractive along all the directions of positive values of the couplings and not just along the eigendirection of the non vanishing eigenvalue.

Similar results exist for region III which is bounded by $N>11$ and $M<\frac{11}{2}N$. Here the first fixed point does not exist while the second fixed point exists and is attractive along a single direction. The third fixed point does not exist anywhere in region III. We plot this in the right plot of Fig. \ref{Flow}.

Region IV is bounded by $N>11$ and $M>\frac{11}{2}N$. Here none of the couplings are asymptotically free.

\section{Sketching out phenomenological implications}
\label{pheno}
To better elucidate the plethora of interesting critical phenomena that emerge from our analysis when considering semi-simple gauge groups we first provide the actual runnings of the gauge couplings for selected interesting renormalisation group trajectories. We then motivate how the unveiled phenomena can spur new ideas for phenomenological applications or be embedded within earlier paradigms.

\begin{figure}[htbp]
\center
\includegraphics[width=0.4\textwidth]{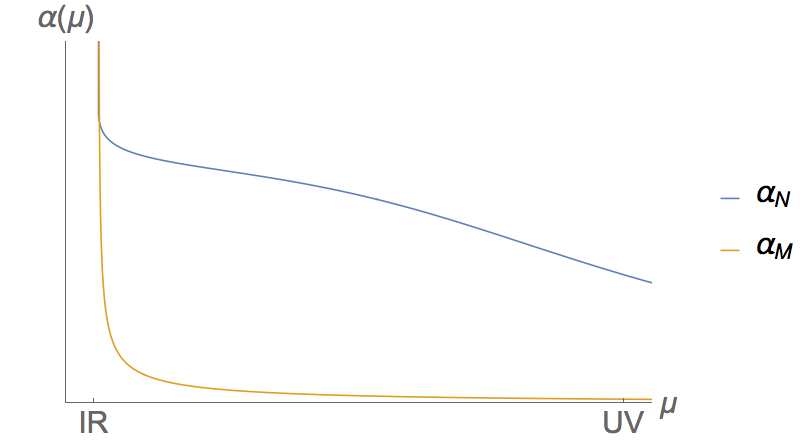}
\caption{The running of $\alpha_N$ and $\alpha_M$ in the $SU(N)\times SU(M)_L$ theory for $N=7$ and $M=25$ along a trajectory in the left plot of Fig. \ref{FPflow} close to the fixed point on the $\alpha_N$ axis.}\label{runningNM}
\end{figure}

Let us start with displaying the running  in Fig.~\ref{runningNM} for a trajectory very close to the fixed point in the left plot of Fig.~\ref{FPflow}. 
This running shows $\alpha_N$ and $\alpha_M$ starting in the UV as asymptotically free and then approaching (but not reaching) the fixed point on the $\alpha_N$ axis. Here at intermediate scales both couplings run very slowly and are near-conformal, i.e. they walk \cite{Holdom:1984sk,Holdom:1983kw,Holdom:1981rm,Yamawaki:1985zg,Appelquist:1986an} (see \cite{Sannino:2009za} for a review). As the deep IR is approached the couplings again grow to larger values. We have, therefore, uncovered an independent mechanism to generate walking theories that uses the gauging of part of their flavour symmetries. Given that the Standard Model is already a semi-simple gauge group we find this way of constructing walking theories among the most natural ways explored so far in the literature. 

\begin{figure}
\centering
\includegraphics[width=0.6\textwidth]{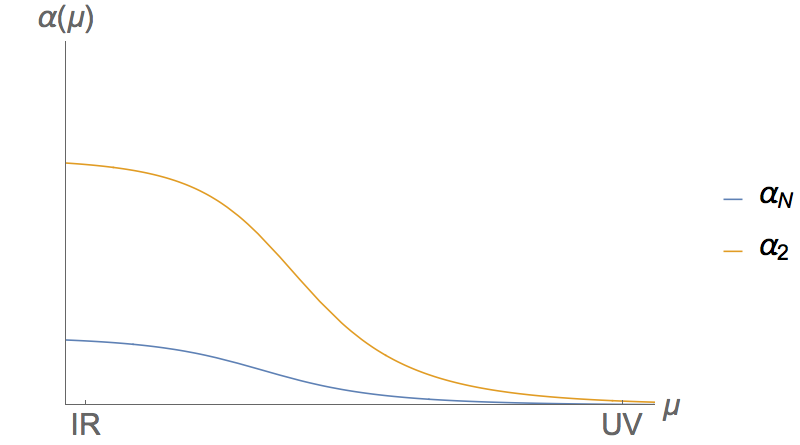}
\caption{The running of the coupling of $\alpha_N$ and $\alpha_2$ in the $SU(N)\times SU(2)_L$ theory for $N=9$ and $M=40$ along a trajectory in the left plot of Fig. \ref{fig:flow13123} from the UV gaussian fixed point to the central infrared interacting fixed point.} \label{N2IR}
\end{figure}

We also show in Fig. \ref{N2IR} the running of the couplings in the $SU(N)\times SU(2)_L$ for $N=9$ and $M=40$ along a trajectory where both couplings are asymptotically free and simultaneously reach the central interacting IR fixed point of the left plot in Fig. \ref{fig:flow13123}.

\begin{figure}[ht]
\begin{minipage}[b]{0.45\linewidth}
\centering
\includegraphics[width=\textwidth]{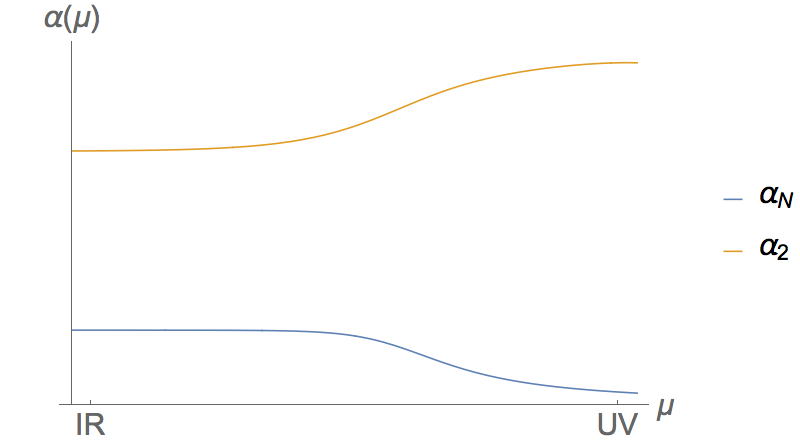}
\end{minipage}
\hspace{0.5cm}
\begin{minipage}[b]{0.45\linewidth}
\centering
\includegraphics[width=\textwidth]{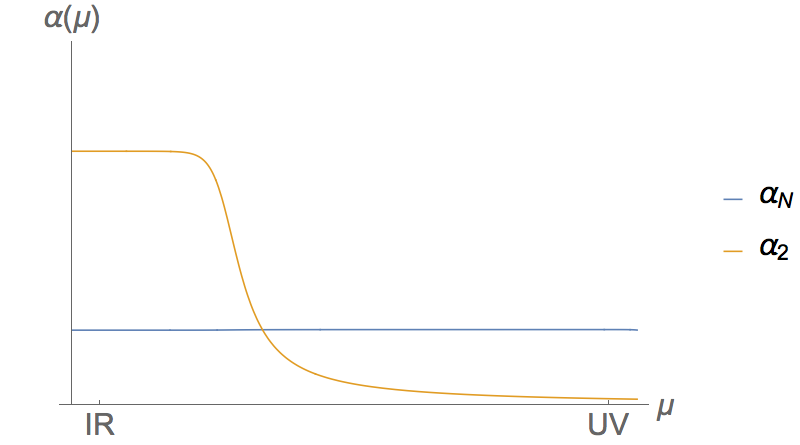}
\end{minipage}
\caption{The left plot shows, the running of $\alpha_N$ and $\alpha_2$ in the $SU(N)\times SU(2)_L$ theory for $N=9$ and $M=40$ along the fine tuned safety free trajectory from the central infrared fixed point to the ultraviolet fixed point on the $\alpha_2$ axis. The right plot shows, the running of $\alpha_N$ and $\alpha_2$ of the same theory but now along the fine tuned safety free trajectory from the central infrared fixed point to the ultraviolet fixed point on the $\alpha_N$ axis.}
\label{runningN2a}
\end{figure}

We now move on to another interesting scenario shown in Fig.~\ref{runningN2a}. Here we observe the phenomenon of  {\it safety free} behaviour according to which one coupling is asymptotically safe and the other is asymptotically free while in the infrared they both reach the infrared fixed point. This is only possible in the $SU(N) \times SU(2)_L$ gauge theory scenario. This scenario is quite different from the case of either complete asymptotic freedom or safety. Nevertheless the theory is still well defined both in the UV and in the IR, opening the door to new ways of constructing extensions of the Standard Model where some gauge interactions are asymptotically free and others are asymptotically safe. It is worth stressing that these new phases are possible because of the interplay of two gauge sectors. 

\begin{figure}[ht]
\begin{minipage}[b]{0.45\linewidth}
\centering
\includegraphics[width=\textwidth]{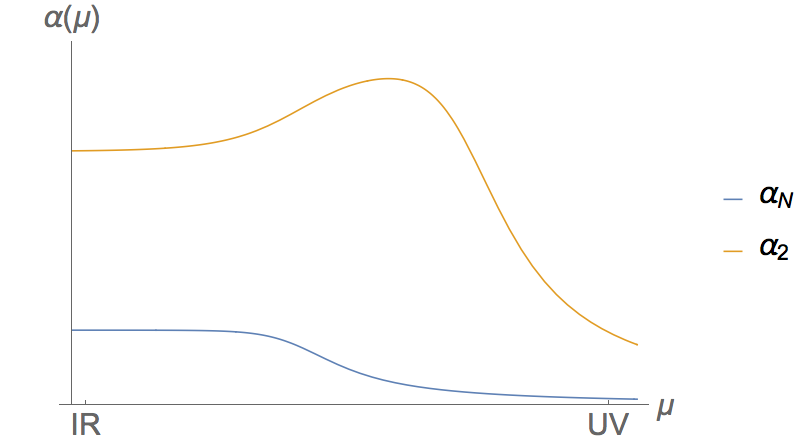}
\end{minipage}
\hspace{0.5cm}
\begin{minipage}[b]{0.45\linewidth}
\centering
\includegraphics[width=\textwidth]{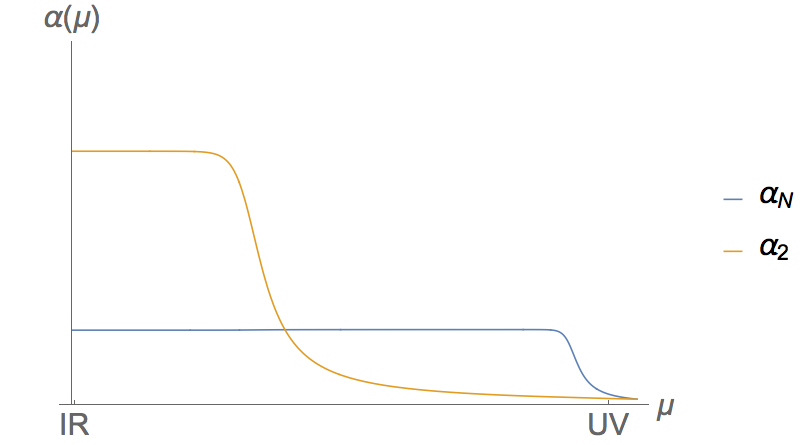}
\end{minipage}
\caption{The left plot shows, the running of $\alpha_N$ and $\alpha_2$ in the $SU(N)\times SU(2)_L$ theory for $N=9$ and $M=40$ along a trajectory in the left plot of Fig. \ref{fig:flow13123}. The flow starts off near the gaussian ultraviolet fixed point and then flows close to the interacting fixed point along the $\alpha_2$ axis but settling at the central infrared interacting fixed point. The right plot shows, the running of $\alpha_N$ and $\alpha_2$ of the same theory but now along a trajectory flowing from the gaussian ultraviolet fixed point, then flowing close to the infrared interacting fixed point a long the $\alpha_N$ axis but settling at the central infrared interacting fixed point.  } 
\label{runningN2b}
\end{figure}

An intriguing behaviour is the one shown in the left panel of  Fig.~\ref{runningN2b}. Here one observes the interaction strength of one of the two couplings reach a maximum at some intermediate energy scale while decreasing both in the UV and the IR. This peculiar behaviour might turn out to be useful when discussing dark matter properties because if it is realised it could help alleviate phenomenological constraints on symmetric dark matter models along the lines suggested in \cite{Sannino:2014lxa}. The right panel of Fig.~\ref{runningN2b} displays a theory where both couplings are asymptotically free and both reach an interacting IR fixed point. 

Although our analysis seems to suggest that the above phases only emerge when both $N$ and $M$ are considerably large (making realistic model building difficult) we stress that our analysis is bounded by perturbation theory. 

Consider supersymmetric QCD with a single $SU(N)$ gauge group and with a variable number of superflavors in the fundamental representation for which the exact conformal window was derived by Seiberg in \cite{Seiberg:1994pq}.  Here the lower bound of the conformal window extends well below the boundary set by the value of the coupling constant at the two loop infrared fixed point reaching unity \cite{Ryttov:2007sr}. In fact the conformal window extends just below the point where the coupling constant at two loops has become arbitrarily large and where one would have naively guessed that no infrared fixed points could have existed. If this is a generic feature then the regions in Fig. \ref{fig:n2} bounded by the solid black and orange curves become larger. Specifically the solid orange lines are shifted towards smaller values of $N$ and $M$. In other words taking into account non-perturbative effects is likely to enlarge the regions in which the above phases exist proving a model template which is more realistic. 

Lastly we point out that it is likely that once scalars are included the emerging phase diagram will be more involved. For instance the UV fixed point uncovered in \cite{Litim:2014uca} for a specific simple gauge-Yukawa theory can only be seen once the effects from all the couplings in the theory are included. To study the inclusion of Yukawa and scalar self couplings in a semi-simple gauge-Yukawa theory is the next appropriate step we believe one should take. 
 
\section{Conclusions}
\label{conclusions}
We investigated the quantum critical behaviour of relevant classes of semi-simple fermionic gauge theories resembling the Standard Model.  In particular we studied an $SU(N)$ gauge theory with $M$ Dirac flavors where we first gauged an $SU(M)_L$ and then an $SU(2)_L \subset SU(M)_L$ of the original global symmetry $SU(M)_L\times SU(M)_R \times U(1) $ of the theory. Lepton-like particles were added to   avoid gauge anomalies. We showed that at the two-loops level an intriguing phase diagram appears. New phases emerged in which one, two or three fixed points have been shown to exist.  We also unveiled a phase featuring complete asymptotic freedom and simultaneously an interacting infrared fixed point in both couplings. Intriguingly the analysis further revealed the existence of special renormalisation group trajectories along which one coupling displays asymptotic freedom and the other asymptotic safety, while both flow in the infrared to an interacting fixed point. These  are the "safety free" trajectories. We further discussed the associated renormalisation group flow of the coupling constants. The knowledge of the quantum critical behaviour of these types of theories is useful information when constructing beyond Standard Model scenarios. For instance we discovered a new genuine way of producing near-conformal (i.e. walking) theories. This interesting scenario emerges due to a nontrivial interplay between the different gauge couplings and is seen specifically for semi-simple gauge theories.

\acknowledgments 
We thank Esben M\o lgaard for discussions. This work is partially supported by the Danish National Research Foundation grant DNRF:90. 


\appendix

\section{Beta Functions Coefficients for $SU(N)\times SU(M)_L$}\label{app:NM}
 
We here provide the beta function coefficients for the $SU(N)\times SU(M)_L$ theory studied in Section \ref{sec:NM} up to two loops
\begin{eqnarray}
a_N&=&\frac{11}{3}N-\frac{2}{3}M \ ,\\
b_N&=&\frac{17}{3}N^2-\frac{5}{3}NM-\left(\frac{N^2-1}{2N}\right)M \ ,\\
c_N&=& -\frac{1}{4}\left(M^2-1\right) \ ,  \\
a_M&=&\frac{11}{3}M-\frac{2}{3}N \  ,\\
b_M&=& \frac{17}{3}M^2-\frac{5}{3}NM-\left(\frac{M^2-1}{2M}\right)N \ ,\\
c_M&=&- \frac{1}{4}\left({N^2-1}\right) \ .
\end{eqnarray}

\section{Fixed Points in $SU(N)\times SU(M)_L$}\label{app:fixed}

Here we show that the third fixed point FP$_3$ does not exist for the $SU(N)\times SU(M)_L$ theory. Consider first the one-loop coefficient for a simple gauge group theory and denote it by $a$. We will imagine tuning the number of Weyl fermions $N_w$ around the critical point where asymptotic freedom is lost/gained by parametrising small departures away from $a=0$ via
\beq
N_w = \frac{11}{2} \frac{C_2(G)}{T(r)} - \epsilon
\eeq
The choice $\epsilon=0$ corresponds to $a=0$. If $\epsilon > 0$  the theory is asymptotically free and if $\epsilon < 0$ the theory is infrared free. Here we assume that $N_w$ can take continues values which then implies that $\epsilon$ is a continuous parameter. This is only for pedagogical reasons and we will later use integer numbers for the matter fields.

Now we write the two-loop coefficient b in terms of the parameter $\epsilon$
\beq
b=-\frac{7}{2}C_2(G)^2-\frac{11}{2}C_2(G)C_2(r)+\left( \frac{5}{3}C_2(G) +C_2(r) \right)T(r) \epsilon\;.
\eeq
We observe that for an infrared free theory, where $\epsilon < 0$, the two-loop coefficient will always be negative. On the other hand for an asymptotically free theory, where $\epsilon > 0$, there exists a region in which $b$ is also negative. This can happen since $\epsilon$ can be tuned arbitrary small. In this limit the system then has a fixed point which is the usual Banks-Zaks fixed point.

We will now proceed to discuss the $SU(N)\times SU(M)$ theory. Here the fixed point, FP$_3$, induced by the mixing of the semi-simple gauge group structure is in terms of the beta function coefficients given by,
\beq
\alpha_{N}^* = - 2\pi \frac{a_Nb_M - a_Mc_N}{b_Nb_M - c_Nc_M} \;,\qquad \alpha_{M}^* = -2\pi \frac{a_Mb_N - a_N c_M}{b_Nb_M - c_Nc_M}\;.
\eeq
Lastly we observe that the mixing coefficients $c_N$ and $c_M$ are always negative. 

We will start by discussing the case of region II and III where either $SU(N)$ or $SU(M)$ is infrared free and then proceed to region I where both groups are asymptotically free.

\subsection*{Region II and III}
In these two regions we have either $a_N < 0$ and $a_M > 0$ (region II) or $a_N > 0$ and $a_M < 0$ (region III). 

We will first consider region II. From the above considerations we know that $b_N < 0$ since $a_N < 0$ and that $b_M$ can be either positive of negative. Furthermore $c_N < 0$ and $c_M < 0$, thus the fixed point FP$_3$ can be written as

\beq
\alpha_{N}^* = - 2\pi \frac{ -\vert a_N \vert b_M + \vert a_Mc_N \vert}{-\vert b_N\vert b_M -\vert c_Nc_M \vert} \;,\qquad \alpha_{M}^* = -2\pi \frac{- \vert a_M  b_N \vert  - \vert a_N c_M \vert }{-\vert b_N \vert b_M - \vert c_Nc_M \vert }\;.
\eeq
Recall that we require $\alpha_{N}^* > 0$ and $\alpha_{M}^* > 0$ for the fixed point to be physical. We now observe that if $b_M \geq 0$ then $\alpha_{M}^*$ will be overall negative and thus non-physical. If $b_M < 0$ then  for $\alpha_{N}^*$ to be positive $\vert b_Nb_M \vert < \vert c_Nc_M \vert $ has to hold while for $\alpha_{M}^*$ to be positive $\vert b_Nb_M \vert > \vert c_Nc_M \vert$ must be satisfied. This clearly can never happen and the fixed point cannot exist. By the same logic we can reach a similar conclusion for region III. It should then be clear that FP$_3$ fails to exist in region II and III.

\subsection*{Region I}
In region I we have $a_N > 0$ and $a_M > 0$ together with $c_N < 0$ and $c_M < 0$. Whether FP$_3$ is physical depends then on the sign of the remaining coefficients $b_N$ and $b_M$. Since both $b_N$ and $b_M$ can be either positive or negative our first task is to find for what specific values of $N$ and $M$ they change sign. Both coefficients are given in Appendix \ref{app:NM} and the sign is dictated by
\begin{align}
b_N< 0 : & \qquad \frac{34N^2}{13N^2-3} < \frac{M}{N} < \frac{11}{2}  \;,\\
b_N> 0 : & \qquad \frac{34N^2}{13N^2-3} >  \frac{M}{N} > \frac{2}{11}  \;,\\
b_M< 0:& \qquad  \left( \frac{34M^2}{13M^2-3} \right)^{-1} > \frac{M}{N} > \frac{2}{11}  \;. \\
b_M>0: & \qquad \left( \frac{34M^2}{13M^2-3} \right)^{-1} < \frac{M}{N} < \frac{11}{2}
\end{align}
First it is easy to check that the situtation with $b_N<0$ and $b_M<0$ cannot be realised. Hence there are in principle three possibilities for the fixed point to be physical: If $b_N>0$ and $b_M<0$ then $\vert a_Mc_N\vert> \vert a_Nb_M \vert$  must be satisfied, if $b_N>0$ and $b_M>0$ then $\vert c_Nc_M \vert>\vert b_Nb_M\vert$ must be satisfied and if $b_N<0$ and $b_M>0$ then $\vert a_Nc_M \vert>\vert a_Mb_N\vert$ must be satisfied. However one can check by explicit use of the beta function coefficients that this can never occur. Hence the third fixed point solution FP$_3$ does not exist in region I, II and III.

\section{Beta Functions Coefficients for $SU(N)\times SU(2)_L$}\label{app:N2L}

We here provide the beta function coefficients for the $SU(N)\times SU(2)_L$ theory studied in Section \ref{sec:N2} up to two loops
\begin{eqnarray}
a_{N} &=& \frac{11}{3}N -\frac{2}{3}M \ , \\
b_{N} &=& \frac{17}{3}N^2 - \left( \frac{5}{3}N + \frac{N^2-1}{2N} \right) M \ ,  \\
c_{N} &=& - \frac{3}{4} \ , \\
a_{2} &=& \frac{22}{3} - \frac{2}{3}N \ ,  \\
b_{2} &=& \frac{68}{3} - \frac{49}{12}N \ ,  \\
c_{2} &=& \frac{1-N^2}{4} \ . 
\end{eqnarray}

\end{document}